# Note on: "Inevitability of Plate Tectonics on Super-Earths" by Valencia, O'Connell and Sasselov, arXiv preprint 0710.0699

Mensur Omerbashich

Eötvös Loránd Geophysics Institute of Hungary, Columbus u.17-23, Budapest
Ph. +36-1-252-4999 x151, Fax +36-1-363-7256, omerbashich@gmail.com

Valencia *et al.* recently claimed that the mass of a Super-Earth (SE) is a sole factor in determining whether a SE is tectonically active or not. However, mass-resolving astrometry is unable to discern between a SE and its moons if any. The fact that no exomoons have been discovered yet is rather a matter of instrumentation imperfection at the present, not of physical absence of exomoons. This, with recently discovered relationships between geometric and physical properties in astronomical bodies (Transiting planets; the Earth) makes it impossible to know yet if the Wagener's (here constraining) supposition on somehow-tidally caused tectonics holds universally or not also.

[1] recently claimed that it is a mass of a Super-Earth (SE) alone, and not the external (here lunar) forces, that can be one-on-one associated with SE tectonics. However, one can estimate the mass of a SE only generally, i.e. without being able to discern whether the masses of one or more of its natural satellites are also included in a specific estimate or not. Then there is no basis for *a priori* dismissing the Wagener's fundamental proposal on tidal friction as being (somehow) responsible for tectonogenesis. Moreover, new proposals in the same direction to his have been worked out, such as [2], offering a new approach to the above Wagener's supposition; for instance, [2] has found out that magnification of all Earth masses' (mainly mantle's) lunar-synodic resonance can give rise to Earth tectonics... Furthermore, extending this concept to all scales resulted in a set of remarkable scaling relationships amongst the Newtonian gravitational constant – along the mechanist and quantum scales – with satellite-orbital and grave-mode periods of our planet [3]. This added to a captivating relationship as reported recently for periods v. masses of Transiting planets [4]. Such a series of previously unthinkable relationships between astronomy (geometry) on one side, and astrophysics (mass) on the other, strongly reaffirms the Wagener's original ideas on moon(s)-caused tectonogenesis. At the same time, a vast number of astronomical bodies calls for caution, to say the least, before labeling an arbitrary and externally unconstrained concept as universal.

## Conclusion

By ignoring the possibility for any future discoveries of exomoons, as well as of their known and unknown relationships with exoplanets, the criticized paper had jumped to the conclusion that, just by looking at its apparent mass, a Super-Earth can be *a priori* said to be exhibiting (exomoon-independent) tectonics.